\documentstyle[12pt]{article}

\newlength{\extraspace}
\newlength{\extraspaces}
\newcommand{\be}{\begin{equation}}
\newcommand{\ee}{\end{equation}}
\newcommand{\bea}{\begin{eqnarray}}
\newcommand{\nn}{\nonumber}
\newcommand{\eea}{\end{eqnarray}}

\newcommand{\nk}{\noindent}

\newcommand{\xd}{\chi}

\newcommand{\yy}{W}

\def\lsim{\mathrel{\rlap {\raise.5ex\hbox{$ < $}}
{\lower.5ex\hbox{$\sim$}}}}

\def\gappeq{\mathrel{\rlap {\raise.5ex\hbox{$>$}}
{\lower.5ex\hbox{$\sim$}}}}
\def\lappeq{\mathrel{\rlap{\raise.5ex\hbox{$<$}}
{\lower.5ex\hbox{$\sim$}}}}

\begin{document}

\begin{titlepage}
\begin{flushright}
OUTP-97-52P \\
hep-th/9711006\\
September 1997\\
\end{flushright}
\begin{centering}
\vspace{.1in}
{\large {\bf T duality for boundary-non-critical strings}}\\
\vspace{.4in}
{\bf G. Amelino-Camelia} and
{\bf N.E. Mavromatos$^{\diamond}$}\\
%$^{b,\diamond}$}\\
%
\vspace{.03in}
\nk 
%$^a$ 
University of Oxford, Theoretical Physics,
1 Keble Road, Oxford OX1 3NP, U.K.   \\

\vspace{1in}
{\bf Abstract} \\
\vspace{.05in}
\end{centering}
{\small  Recent work on the action of T duality on Dirichlet-branes
is generalized to the case in which the open string satisfies
boundary conditions that are neither Neumann nor Dirichlet.
This is achieved by implementing T duality as a canonical transformation
of the $\sigma$-model path integral.
A class of boundary interactions that violate conformal symmetry
is found to be T-dual of a correspondingly non-conformal class
of boundary conditions. 
The analogy with some problems in
boundary-non-critical quantum mechanics of interest for 
condensed matter is pointed out.}
\vspace{0.2in}
\vspace{0.01in}
\vfill
$^{\diamond}$ P.P.A.R.C. Advanced Fellow.
\end{titlepage}
\newpage

%\section{Introduction}

A series of dramatic break-throughs has recently brought
our understanding of critical open strings to levels unimaginable 
even just a few years ago~\cite{dbranes}.
Part of the motivation for 
the analysis reported in this Letter
comes from contemplating the possibility
that some of the technical tools
that have contributed to the recent progress
in the study of critical open strings
might be generalized also to non-critical open strings,
thereby providing an opportunity 
for new insight in this class of theories.

We are here particularly interested in open string theories
in which the violations of conformal invariance
originate from the physics of the boundary of the string.
It is in fact plausible that these theories
turn out to be more manageable than their bulk-non-critical
counterparts, while still providing a reach laboratory 
for the exploration of exciting issues, such as the fate of
the short-distance structure of the target-space geometry
when world-sheet conformal invariance is violated~\cite{emn,aemn}.
While we were writing this report we became aware of Ref.~\cite{new1},
whose findings appear to suggest that
boundary-non-critical strings might also have a place in very
conventional string-theory frameworks. 

The expectation of the existence of ``reasonably treatable''
boundary-non-critical strings finds support in the established 
literature on boundary-conformal field theory~\cite{bcft,gacdelta,gacBCQM},
which has provided examples of boundary-non-critical deformations
that preserve (most of) the simplicity ({\it e.g.} integrability)
of the original conformal-invariant theory.
Of course, a difficult task in the development of such ``reasonably 
treatable'' boundary-non-critical strings 
is the identification of non-conformal boundary deformations
that be respectful (as much as possible) of the simplicity 
of critical open strings.
In this letter we provide no answer to this 
most challenging aspect of the problem at hand; in fact,
we limit ourself to the study of a specific type
of boundary deformation: a linear-dilaton boundary background.
While we find no reason to believe that this particular deformation
should eventually turn out to be particularly simple and treatable,
its analysis is intrinsically well motivated in light of
the Liouville approach to target time in the context of 
string theory~\cite{emn}, or in the context of certain cosmological models
based on strings~\cite{aben}. 

The linear-dilaton boundary backgrounds considered here
have the merit of providing us a concrete framework
in which to start exploring the applicability to 
boundary-non-critical strings of T duality, which is
one of the most important 
tools in the derivation of nonperturbative results
of critical open strings.
This is the central point of this Letter.
We find that a conventional path-integral formalism \footnote{We view
the T duality as a canonical transformation in a $\sigma$-model 
path integral.
This point of view was taken in 
the original literature on Kramers-Wannier 
duality in statistical physics, and recently, 
in the context of critical string
theory, in ref. \cite{dornotto,kow}.}
for the implementation of T duality transformations
also applies to the boundary-non-critical case.
The implications of T duality in this context
are such that our linear-dilaton boundary backgrounds 
are T-dual of a correspondingly non-conformal class
of boundary conditions. 
We report these observations as intuitively
as possible via the analysis of two explicit cases,
rather than providing general and formal derivations.
The two cases considered are 
the flat 26-dimensional target space time
and the 2-dimensional ``cigar'' target space time~\cite{witt}.
Additional intuition for our results can be gained by 
considering corresponding results in deformed boundary-conformal
field theory and quantum mechanics. 
For example, among the
recent results on two-anyon quantum mechanics one 
finds the observation \cite{gacBCQM,gacbak} that 
this scale-invariant system can be 
equivalently deformed either by a contact 
interaction $\delta^{(2)}({\bf r})$ with running
coupling $g(\mu)$ or by enforcing in the s-wave sector
scale-dependent boundary conditions of the type
\begin{equation}
\left[r^{|\nu|} \psi({\bf r}) - w \rho^{2 |\nu|}
{d\left(r^{|\nu|} \psi ({\bf r})\right)
\over d (r^{2 |\nu|})}\right]_{r=0}=0 ~,
\label{bc1}
\end{equation}
where ${\bf r}$ is relative-position vector,
$r \equiv |{\bf r}|$, $\nu$ is the ``statistical parameter'' 
(a fundamental parameter of anyonic physics),
while $\rho$ is a reference scale which together with 
the dimensionless parameter $w$
can be put in correspondence with the $g(\mu)$ characterizing
the contact-interaction formulation of the problem.

\vskip.2in

\noindent
{\bf A. Flat 26-dimensional target space time.}
Consider the case of a simple linear dilaton boundary 
background~\cite{aben},
in a non-critical string theory 
with central charge deficit $Q^2$ 
\be
S= \frac{1}{4\pi {}}\int d^2 \sigma \partial X^i {\overline \partial} X^i
+ \frac{1}{4\pi {}}\int d^2 \sigma \partial Y^j {\overline \partial} Y^j 
- \int_{\partial \Sigma } Q {\hat k} \eta^i X^i 
\label{part}
\ee
where $\eta$ is an n-dimensional constant number-valued vector, and
(for later convenience) we have divided the 26 fields into $n$ fields
of type $X$ ({\it i.e.} $i = 1 \dots n$) and $26-n$  fields of type $Y$
({\it i.e.}  $j = 1 \dots 26 -n$). We have also used
conventional notations ${\hat k}$ for the extrinsic curvature 
of the fiducial metric, 
$\Sigma$ for the world-sheet manifold, and $\partial \Sigma$
for the boundary of the world-sheet manifold.
The fields $X$ and $Y$ are assumed to satisfy Neumann boundary 
conditions: 
\be
  \partial_\sigma \hat X^i =  \partial _\sigma \hat Y^j =0 
\qquad {\rm on~}\partial \Sigma
\label{neuman}
\ee
where we use the notation $\partial_\sigma$ ($\partial_\tau$) 
for normal (tangent) derivatives\footnote{Consistently with 
the objective of keeping our discussion as intuitive as possible
we shall often implicitly or explicitly
assume the fiducial geometry to be that of a disk, 
with constant extrinsic
curvature (e.g. ${\hat k}=2$ for a unit disk).}
on $\partial \Sigma$
and we also use the notation $\hat \Phi$ to emphasize that
a field $\Phi$ is being evaluated on  $\partial \Sigma$. 

The path-integral formulation of this $\sigma$-model is: 
\be
Z=\int DX \, \delta (\partial_\sigma \hat X^i)
\, \delta (\partial_\sigma \hat Y^j) \,
\exp \left[ -\int _{\Sigma} \frac{1}{4\pi{}} 
\left[ \partial X^i {\overline \partial} X^i +
\partial Y^j {\overline \partial} Y^j \right]
+ \int _{\partial \Sigma} {\hat k} Q \eta^i X^i \right]
~,
\label{opensigma}
\ee
where we indicated explicitly via a boundary delta-functional
that the $X^i$ and the $Y^j$ are Neumann fields.

We wish to apply a functional T duality transformation on the fields $X^i$
in the path integral (\ref{opensigma}).  
We generalize recent work on T duality in the path-integral formalism
(see {\it e.g.} Ref.~\cite{dornotto}), which has been exclusively concerned 
with open strings satisfying 
conformal (Dirichlet or Neumann) boundary conditions,
to the simple class of non-conformal boundary conditions that correspond
to the boundary interaction described by our
linear dilaton boundary background.
We find that the same procedure of T-dualization
is appropriate in the case of nonconformal boundary conditions.
In particular, the T duality transformation 
has again (compare with Ref.~\cite{dornotto}) 
as crucial element the introduction of a vectorial field variable
corresponding to the partial derivative of the fields $X^i$
that are being dualized:
\be
\yy^i_\alpha \equiv \partial_\alpha X^i ~.
\label{redef}
\ee
The fields $\yy$ are introduced in the path integral
via the identity:
\be
\int D\yy^i_\alpha \, \delta (\yy^i_\alpha-\partial_\alpha X)
\, \delta (\epsilon_{\alpha\beta}\partial_\alpha \yy^i_\beta)
=1 ~,
\label{constr}
\ee
which takes of course into account 
the `Bianchi identity':
\be
\epsilon_{\alpha\beta}\partial_\alpha \yy_\beta=0
\label{bianchi}
\ee
The path integral (\ref{opensigma}) is therefore rewritten as:
\bea
&~& Z= \int DX \, DY \, D\yy  
\, \delta (\partial_\sigma \hat X^i) \,
\delta (\partial_\sigma \hat Y^j) \,
\delta (\yy^i_\alpha-\partial_\alpha X^i)
\, \delta (\epsilon_{\alpha\beta}\partial_\alpha \yy^i_\beta) \,
\delta (\hat \yy^i_\sigma -\partial_\sigma \hat X^i)  
\nn \\
&~&~~~~~~~~~~~~
\delta (\hat \yy^i_\tau -\partial_\tau \hat X^i) \,
\exp \left[
-\frac{1}{4\pi {}}\int_\Sigma (\yy^i_\alpha)^2 
-\frac{1}{4\pi {}}\int_\Sigma \partial Y^j {\overline \partial} Y^j 
+ \int _{\partial \Sigma} {\hat k} Q \eta^i X^i \right] 
\nn \\  
&~&~~~= Z_Y \int DX^i \, D\yy^i_\alpha \, D\xd^i \, 
D\lambda^i_\alpha \, 
\delta (\partial_\sigma \hat X^i) 
\nn \\
&~&~~~~~~~~~~~~
\exp \left[
-\frac{1}{4\pi {}} \int_\Sigma  (\yy^i_\alpha)^2 
- i \int_\Sigma \xd^i (\epsilon_{\alpha\beta}\partial_\alpha \yy^i_\beta) 
- i \int_\Sigma \lambda^i_\alpha (\yy^i_\alpha-\partial_\alpha X) \right] 
\nn \\  
&~&~~~~~~~~~~~~
\exp \left[ \int_{\partial \Sigma} {\hat k} Q \eta^i X^i 
- i \int_{\partial \Sigma} \hat \lambda^i_\sigma \hat \yy^i_\sigma 
- i \int_{\partial \Sigma} \hat \lambda^i_\tau 
(\hat \yy^i_\tau -\partial_\tau \hat X^i) \right] 
\label{prepixtilde}
\eea
where (consistently with the notation already introduced for the normal
and tangent derivatives) we denoted the normal (tangent) components 
of world-sheet vectors with a lower index $\sigma$ ($\tau$). 
(Summation is of course understood on all repeated indices apart 
from $\sigma$ and $\tau$ which are fixed labels for boundary fields.)
We also introduced the short-hand notation
\bea
&~& Z_Y = \int DY \,
\delta (\partial_\sigma \hat Y^j) \,
\exp \left[
-\frac{1}{4\pi {}}\int_\Sigma \partial Y^j {\overline \partial} Y^j 
\right] 
\label{zy}
\eea
for the portion of the partition function that concerns
the $Y^j$ fields, which are ``spectators'' of the T duality
transformation being performed on the $X^i$ degrees of freedom;
moreover we adopt the convention $\epsilon_{\sigma\tau}=1$
and the following functional representation of a $\delta (\phi)$ constraint
\be
 \delta (\phi) = \int D\lambda e^{-i\int _{M} \phi \lambda}
\label{deltconstrfr}
\ee
with $M$ an appropriate manifold, indicating the range of definition of 
the arguments of the fields $\phi$. 
In our case $M=\Sigma$ or $\partial \Sigma$.

The Lagrange multipliers fields 
$\xd^i$ and $\lambda^i_\alpha$ play a highly non-trivial
r\^ole in the T duality
transformation; in particular, the fields $\xd^i$,
which implement the Bianchi identity (\ref{bianchi}), 
turn out to be
directly related to the fields that are T-dual to the fields $X^i$,
just as expected from the analysis reported in Ref.~\cite{dornotto}. 

It is convenient to rewrite (also using integration by parts) 
the partition function of (\ref{prepixtilde}) as
\bea
&~& Z= Z_Y \int DX^i \, D\yy^i_\alpha \, D\xd^i \, D\lambda^i_\alpha
\delta (\partial_\sigma \hat X^i) 
\nn \\
&~&~~~~~~~~~~~~
\exp \left[ \int_\Sigma  \left( \frac{i}{2 \sqrt{\pi {}}} \yy^i_\alpha 
+ \sqrt{\pi {}} \left( \epsilon_{\alpha \beta} \partial_\beta \xd^i
- \lambda^i_\alpha \right) \right)^2 
- \pi {}  \int_\Sigma  \left( \epsilon_{\alpha \beta} 
\partial_\beta \xd^i - \lambda^i_\alpha \right)^2 \right]
\nn \\  
&~&~~~~~~~~~~~~
\exp \left[ i \int_\Sigma X^i \partial_\alpha \lambda^i_\alpha 
- i \int_{\partial \Sigma} \hat X^i \left( \hat \lambda^i_\sigma 
+ \partial_\tau \hat X^i
+ i {\hat k} Q \eta^i \right)
\right]
\nn \\  
&~&~~~~~~~~~~~~
\exp \left[ 
- i \int_{\partial \Sigma} \hat \yy^i_\tau \, (\hat \lambda^i_\tau - \xd^i)
- i \int_{\partial \Sigma} \hat \yy^i_\sigma \, \hat \lambda^i_\sigma \right]
\label{pixtilde}
\eea

The functional integration over $X$ and $\yy$ can be done easily;
one obtains (up to an irrelevant overall factor coming from the 
gaussian integration over $\yy$)

\bea 
&~&Z= Z_Y \int D \xd \, D\lambda \, \delta (\partial_\alpha \lambda_\alpha)
\, \delta(\hat \lambda_\sigma) \, \delta (\hat \xd  - \hat \lambda_\tau) \,
\delta(\partial_\tau \hat\lambda_\tau + i Q {\hat k} \eta^i 
+ \hat \lambda_\sigma) 
\nn \\  
&~&~~~~~~~~~~~~
\exp \left[-\pi {} \int_{\Sigma} (\lambda^i_\alpha 
- \epsilon_{\alpha\beta}\partial_\beta \xd^i)^2 \right] 
\label{pixtilde2}
\eea

The fields $X_D^i$ that are T-dual to the fields $X^i$
are easily identified as the ones satisfying the relation
\be
\frac{i}{2\pi{}} \epsilon_{\alpha \beta} \partial_\beta X^i_D \equiv 
\epsilon_{\alpha \beta} \partial_\beta \xd^i - \lambda^i_\alpha 
\label{xdualdef}
\ee
whose consistency follows from the 
constraint $\partial_\alpha \lambda^i_\alpha = 0 $ 
(see Eq.~(\ref{pixtilde2})).

Upon the change of variables $\xd^i \rightarrow X^i_D$,
and disposing of the then trivial functional integration over 
the $\lambda$ fields,
one can easily rewrite 
the partition function of (\ref{pixtilde2}) as
(up to another irrelevant overall factor)
\bea 
Z= Z_Y \int D X_D \, \delta(\partial_\tau \hat X^i_D + 2 \pi {}
Q {\hat k} \eta^i ) 
\exp \left[ - \int_{\Sigma} (\partial_\alpha X_D^i)^2 \right] 
\label{final1}
\eea
where we also used the fact that (\ref{xdualdef}), when combined to the
constraint $\hat \lambda^i_\sigma = 0$
(see Eq.~(\ref{pixtilde2})), implies 
$\partial_\tau \hat X^i_D = - i 2 \pi {} \partial_\tau \hat \xd^i$.

The T duality transformation implemented via the path-integral
manipulations that take from (\ref{opensigma}) to (\ref{final1})
evidently maps a Neumann open string with
boundary interactions corresponding to the
linear dilaton boundary background
present in (\ref{opensigma})
into a \underline{free} open string satisfying nonconformal
boundary conditions~\cite{emndecoh}
\bea 
\partial_\tau \hat X^i_D = - 2 \pi {} Q {\hat k} \eta^i 
~.
\label{BCfinal1}
\eea
This boundary condition  
reduces to the Dirichlet boundary condition in the limit $Q \rightarrow 0$.
For every $Q \ne 0$ it encodes a ``conformal anomaly'' 
for the free T-dual theory
that reflects the 
conformal anomaly of the
corresponding boundary interactions
of the original Neumann theory.~\footnote{It would be interesting 
to explore the relation of these findings
with the recent observation~\cite{li}
that fixed Dirichlet boundary conditions are not conformal invariant
in the presence of a linear-dilaton bulk background~\cite{dbranes}.}
As anticipated in the opening of the present Letter,
this is reminiscent of certain dualities encountered
in boundary conformal field theory~\cite{bcft,gacdelta,gacBCQM,gacbak},
where one also maps a problem with nontrivial boundary interactions
into a problem subject to nontrivial boundary 
conditions.~\footnote{The representation, 
via canonical duality transformation, 
as a peculiar boundary condition, opens 
up the way of treating, in a simpler way, more complicated situations, 
like the Liouville boundary dynamics~\cite{ambjorn}.  There, the 
complications arise from
the fact that the Liouville field on the boundary 
does not behave like a scalar field,
in contrast to the linear dilaton case
considered above. It would be interesting to derive 
the corrections to the boundary condition (\ref{BCfinal1}),  
that the Liouville field satisfies due to its non-trivial 
quantum dynamics~\cite{ambjorn}, from the $\sigma$-model 
path integral method. This is left for future work.} 

It is interesting to note at this stage that the
above boundary condition 
might also have important implications for the quantization 
of the central charge.
Consider for instance the boundary condition (\ref{BCfinal1})
in a cordino-disc 
planar geometry (annulus) with a flux passing through 
the middle of the two-dimensional surface. In this case, 
if we parametrize the external boundary $S^1$ 
of the disc by an angular variable 
$0 \le \tau \le 2 \pi$,
and assume for simplicity 
a linear dilaton along one direction only, say ${\hat X^1}$,
{\it assumed to be compact},  
then, the {\it dual} field ${\hat X^1}_D$ may 
have non-trivial winding number $n$ around $S^1$,
${\hat X}^1_D(2\pi)={\hat X}^1_D(0) + 2\pi n R_D$, where $R_D$ 
is a compactification 
radius for the dual field. 
In that case, (\ref{BCfinal1}) 
could be integrated to yield
\be
       2\pi n R_D = - 2\pi {} Q \int d\tau {\hat k} = -4\pi^2 {} Q
\label{euler}
\ee
taking into account the Euler characteristic of a disc to be one. 
The result is a quantization 
of the square-root $Q$ of the 
central charge deficit: 
\be 
     Q = -\frac{n R_D}{2\pi {}} \qquad n \in Z 
\label{conformaldiscr}
\ee
This is consistent with the nature of the boundary non-conformal 
linear dilaton deformation which we introduced to begin with, 
(\ref{opensigma}), provided that the compactification radius of the 
field $i{\hat X}^1$  
is $R_D^{-1}$, as required by T duality ($R = R_D^{-1}$), which is 
consistently reproduced in 
our approach here. Indeed, assuming the quantization condition
(\ref{conformaldiscr}) one finds that
under a shift ${\hat X}^1(2\pi)={\hat X}^1(0) + 2\pi n R_D^{-1}$
the action (\ref{opensigma}) 
changes by factors of $2i\pi m$ (with $m \in Z$)
and thus the partition function remains invariant. 

It would be interesting to explore the relation of the result 
(\ref{conformaldiscr}) with the analogous result
that is found to hold in 
a certain type of (``bulk'') closed strings
with compactified dimensions~\cite{kogancom}. 

\vskip.2in

\noindent
{\bf B. 2-dimensional ``cigar'' target space time.}

The second example we have chosen in order to illustrate 
the doings of T duality in open-string theories with non-conformal
boundary physics is the
two-dimensional 
black hole with a cigar target-space metric~\cite{witt}.
For the closed-string version of
this $\sigma$-model theory it is known that the 
duality transformations map the 
cigar metric into a funnel~\cite{giveon}.
We shall rederive these results using the path-integral 
formulation of T duality;
moreover, we shall generalize them to the
open-string case and allow again our non-conformal
linear-dilaton background on the boundary of the string.

{}From a physical viewpoint,
the open-string analysis of this problem
is motivated by the
$D$-brane representation of the two-dimensional black hole. 
In fact, as argued in ref. \cite{horava}, since the cigar metric 
has two asymptotically flat domains in target space time,
one can get rid of one of them by orbifoldizing the solution 
to obtain a theory of unoriented open and closed
strings in a black hole background, 
and a time-like orbifold singularity 
at the origin. All of the open string states of the model
are confined to the orbifold singularity. 
As discussed in ref. \cite{horava} 
the model is dual, under target-space duality, to
a conventional open string theory in the black hole geometry. 
This was the first example of a $1-$brane (Dirichlet), 
in the context of two-dimensional strings. 

Motivated by these considerations we shall discuss below 
the effect of our T duality canonical transformation,
described above, on the path integral of a $\sigma$-model 
with a cigar-type metric and some open string excitations, 
represented by the existence of a boundary on the corresponding 
$\sigma$-model. Deviations from conformal invariance on the boundary
may for example arise from recoil (or other
matter back-reaction effects) of the Dirichlet brane during 
scattering with some 
string matter~\footnote{In this context, 
it should be noted that 
the duality transformation of ref. \cite{dornotto}
has been applied earlier in ref. \cite{emndecoh} in order
to discuss the formal 
appearance, in a $\sigma$-model framework,  
of Dirichlet boundary conditions
from the vacuum  
in a quantum foam picture of a generic $D$-brane system. 
A certain class of scale-dependent boundary conditions
(boundary conditions that, like the ones
considered in the present paper, are neither fixed Dirichlet, nor Neumann), 
has then been shown to result from
the appearance of a Liouville mode to describe decoherence 
effects due to (boundary) recoil operators.}.  

Let us briefly first review what is known of the duality 
in the two-dimensional Euclidean stringy Black Hole. 
The standard $\sigma$-model action, obtained from gauging the $SL(2,R)/U(1)$ 
Wess-Zumino-Witten-Novikov (WZWN) conformal field theory action, is:
\be
  S^{WZW}=\frac{k}{4\pi}\int _{\Sigma} d^2 \sigma \, [\partial_\alpha \rho
\partial_\alpha \rho + {\rm tanh}^2\rho \, \partial_\alpha \theta 
\partial_\alpha \theta ]
-\frac{1}{8\pi}\int _\Sigma d^2\sigma \, \Phi (\rho) \, R^{(2)}
\label{wzaction}
\ee
where $k$ is the level of the WZWN coset model~\cite{witt},
$\theta$ is a compact $U(1)$ coordinate (Euclidean time), and 
\be
\Phi (\rho)= {\rm ln}{\rm cosh}^2\rho + {\rm constant}
~.
\label{dilaton}
\ee
The above $\sigma$-model 
describes a two-dimensional string moving in a 
Euclidean black hole space time. The Hawking temperature 
of this black hole is given by the radius of the cigar at infinity:
\be
   \beta = T_H^{-1} = \sqrt{\frac{k}{2}}
\label{hawking}
\ee
Conformal invariance requires $k=9/4$, since in that case the central charge
$c=\frac{3k}{k-2}-1$ becomes $26$~\cite{witt}. 

It is convenient to absorb the constant $\frac{k}{2\pi}$ into a redefinition
of $\rho \rightarrow \sqrt{2/k} \, r$. 
Then the action takes the form (which we shall use from now on): 
\be
  S^{WZW}=\frac{1}{2\pi}\int _{\Sigma} d^2 \sigma \, [\partial_\alpha r 
\partial_\alpha r + \epsilon^{-2} \, {\rm tanh}^2 (\epsilon r ) \,
\partial_\alpha \theta \partial_\alpha \theta]
-\frac{1}{8\pi} \int _\Sigma d^2\sigma \, \Phi (r) \, R^{(2)}
\label{wzaction2}
\ee
where $\epsilon=\sqrt{\frac{2}{k}}$ 
and
\be
\Phi (r)={\rm ln}{\rm cosh}^2\epsilon r + {\rm constant}
\label{dilaton2}
\ee

The dual transformation is given by:
\bea 
 &~&   G_{ab} \rightarrow G^{-1}_{ab} \nn \\
&~& \Phi \rightarrow \Phi + \frac{\chi}{2} {\rm ln}{\rm det}G
\label{duality}
\eea
where $\chi$ is the Euler characteristic of the world-sheet manifold
($\chi=2$ for sphere), 
and in the simplified case of a single compactified coordinate
reduces to the standard T duality $R \rightarrow 1/R$~\cite{giveon}. 
Under the above transformation 
one obtains the dual $\sigma$-model action
\be
  S^{WZW}_D=\frac{1}{2\pi}\int _{\Sigma} d^2 \sigma [\partial_\alpha r 
\partial_\alpha r + \epsilon^2 {\rm coth}^2 (\epsilon r )
\partial_\alpha \theta \partial_\alpha \theta]
-\frac{1}{8\pi}\int _\Sigma d^2\sigma \Phi (r, \theta )R^{(2)}
\label{wzactionD}
\ee
with 
\be
\Phi_D(r) = {\rm ln}[\epsilon^{-2} {\rm sinh}^2\epsilon r] + {\rm constant}
\label{dilatonD}
\ee
The Hawking temperature of this dual model is the inverse of (\ref{hawking}):
\be
   \beta^D=\beta^{-1}=\sqrt{\frac{2}{k}}
\label{dualhawking}
\ee
This inversion of temperature is reminiscent (actually a simplified
version) of the standard
Kramers-Wannier duality in statistical physics (c.f. Ising model),
to which the T duality is related. 

We shall show that these results can be easily rederived
upon applying to the $\theta$ field
the path-integral technique of dualization described above.
This result will actually emerge as a corollary 
of the analysis of the theory (\ref{wzaction2})
deformed by our nonconformal dilaton background
\be
  S^{WZWdef} = \frac{1}{2\pi}\int_{\Sigma} 
d^2 \sigma \, [\partial_\alpha r 
\partial_\alpha r + \epsilon^{-2} \, {\rm tanh}^2 (\epsilon r ) \,
\partial_\alpha \theta \partial_\alpha \theta]
-\frac{1}{8\pi} \int _\Sigma d^2\sigma \, \Phi (r) \, R^{(2)}
+ \int_{\partial \Sigma} {\hat k} Q \theta 
\label{action2}
\ee
Our analysis shall follow quite closely the one 
leading to (\ref{pixtilde2}),
but take into account the fact that the cigar target space
is {\it curved}. In particular, 
this implies that in the measure of integration 
of target-space-coordinate fields
one should introduce the appropriate determinants 
$\sqrt{{\rm det}G}$ of the corresponding target space metric tensor,
as required by target-space diffeomorphism invariance.
The starting partition function is therefore
\bea
&~&Z^{WZWdef} = \int Dr \, D \theta \, \sqrt{{\rm det} G} ~
\delta (\partial_\sigma \hat r)
\, \delta (\partial_\sigma \hat \theta) \,
\nn \\  
&~&~~~~~
\exp \left[ - \frac{1}{2\pi} \int_{\Sigma} 
[\partial_\alpha r 
\partial_\alpha r + \epsilon^{-2} \, {\rm tanh}^2 (\epsilon r ) \,
\partial_\alpha \theta \partial_\alpha \theta]
+ \frac{1}{8\pi} \int_\Sigma \Phi (r) \, R^{(2)}
+ \int_{\partial \Sigma} {\hat k} Q \theta \right]
~,~~~~~~~~~~
\label{opensigma2}
\eea
where again we indicated explicitly via functional Dirac $\delta$'s 
that the $r$ and $\theta$ are Neumann fields.
In complete analogy with (\ref{prepixtilde}) we can rewrite
the partition function $Z^{WZWdef}$ as
\bea
&~& Z= \int Dr \, D \theta \, D \yy \, \sqrt{{\rm det} G} ~
\delta (\partial_\sigma \hat r)
\, \delta (\partial_\sigma \hat \theta) \,
\delta (\yy_\alpha-\partial_\alpha \theta)
\, \delta (\epsilon_{\alpha\beta}\partial_\alpha \yy_\beta) \,
\delta (\hat \yy_\sigma -\partial_\sigma \hat \theta)  
\nn \\
&~&~~~~~~~~~~~~
\delta (\hat \yy_\tau -\partial_\tau \hat \theta)  
\nn \\
&~&~~~~~~~~~~~~
\exp \left[ - \frac{1}{2\pi} \int_{\Sigma} 
[\partial_\alpha r 
\partial_\alpha r + \epsilon^{-2} \, {\rm tanh}^2 (\epsilon r ) \,
(\yy_\alpha)^2]
+ \frac{1}{8\pi} \int_\Sigma \Phi (r) \, R^{(2)}
+ \int_{\partial \Sigma} {\hat k} Q \theta \right]
\nn \\
&~&~~~= \int Dr \, D \theta \, D \yy \, 
D\xd \, 
D\lambda_\alpha \, \sqrt{{\rm det} G} \,
\delta (\partial_\sigma \hat r)
\, \delta (\partial_\sigma \hat \theta) \,
\nn \\
&~&~~~
\exp \left[ 
- i \int_\Sigma \xd (\epsilon_{\alpha\beta}\partial_\alpha \yy_\beta) 
- i \int_\Sigma \lambda_\alpha (\yy_\alpha-\partial_\alpha \theta) 
- i \int_{\partial \Sigma} \hat \lambda_\sigma \hat \yy_\sigma 
- i \int_{\partial \Sigma} \hat \lambda_\tau 
(\hat \yy_\tau -\partial_\tau \hat \theta) \right] 
\nn \\
&~&~~~
\exp \left[ - \frac{1}{2\pi} \int_{\Sigma} 
[\partial_\alpha r 
\partial_\alpha r + \epsilon^{-2} \, {\rm tanh}^2 (\epsilon r ) \,
(\yy_\alpha)^2]  
+ \frac{1}{8\pi} \int_\Sigma \Phi (r) \, R^{(2)}
+ \int_{\partial \Sigma} {\hat k} Q \theta \right]
\label{pithetatilde}
\eea
Following again quite closely the analysis in Eqs.(\ref{pixtilde}) 
and (\ref{pixtilde2}) one can easily show that this partition function
can be rewritten as
\bea 
&~&Z= \int Dr \, D \xd \, D\lambda \, \sqrt{{\rm det} G} \,
\delta (\partial_\alpha \lambda_\alpha)
\, \delta(\hat \lambda_\sigma) \, \delta (\hat \xd  - \hat \lambda_\tau) \,
\delta(\partial_\tau \hat\lambda_\tau + i Q {\hat k} 
+ \hat \lambda_\sigma) 
\nn \\  
&~&~~~~~~~
\exp \left[  - \frac{1}{2\pi} \int_{\Sigma} 
\partial_\alpha r 
\partial_\alpha r 
- {\pi \over 2} \int_{\Sigma} coth^2(\epsilon r) (\lambda_\alpha 
- \epsilon_{\alpha\beta}\partial_\beta \xd)^2 
+ \frac{1}{8\pi} \int_\Sigma \Phi (r) \, R^{(2)}
\right]  ~~~~
\label{pixtilde2bis}
\eea

The field $\theta_D$ that is T-dual to $\theta$
is easily identified as the one satisfying the relation
\be
i \epsilon_{\alpha \beta} \partial_\beta \theta_D \equiv 
\epsilon_{\alpha \beta} \partial_\beta \xd - \lambda_\alpha 
\label{xdualdef2}
\ee
Upon the change of variables $\xd \rightarrow \theta_D$,
and introducing the notation ${\tilde G}$
for the target space metric tensor in the dual 
variables $\{r, \theta_D \}$,
one can easily rewrite 
the partition function of (\ref{pixtilde2bis}) as
\bea 
&~&Z= \int Dr \, D \theta_D \, \sqrt{{\rm det} {\tilde G}} \,
\delta(\partial_\tau \hat\theta_D +  Q {\hat k}) 
\nn \\  
&~&~
\exp \left[ - \frac{1}{2\pi} \int_{\Sigma} 
(\partial_\alpha r )^2
- {\pi \over 2} \int_{\Sigma} coth^2(\epsilon r) 
(\partial_\alpha \theta_D)^2 
+ {1 \over 2} \ln {\det G \over \det {\tilde G}}
+ \frac{1}{8\pi} \int_\Sigma \Phi (r) \, R^{(2)}
\right]  ~~~~~~~
\label{final2}
\eea
This shows that 
the action of T duality on the boundary physics
is completely analogous to the one
discussed above for case of a \underline{flat} (26-dimensional) 
target space; in fact, again
the T duality maps a Neumann open string with
boundary interactions corresponding to our linear dilaton background
present in (\ref{opensigma2}) into a free open string satisfying nonconformal
boundary conditions
\bea 
\partial_\tau \hat \theta_D = - Q {\hat k} 
~.
\label{BCfinal2}
\eea
Of course, the discussion following (\ref{BCfinal1}), on 
the quantization of the central 
charge deficit, applies to this case as well.

Concerning the bulk of the string 
it is explicit in (\ref{final2})
that the expected dual ${\tilde G}$ of the original metric
tensor $G$ acts as target space metric tensor in the dual 
variables $\{r, \theta_D \}$.
Eq.(\ref{final2}) also encodes the expected 
shift of the dilaton under T duality.
Consistently with the fact that the dilaton is 
associated with purely quantum effects in the $\sigma$-model 
path integral~\cite{buscher},
the duality transformation of the dilaton (\ref{duality}) comes
from the measure of integration, 
\be 
\sqrt{{\rm det}G/{\rm det}{\tilde G}}
= \exp \left({(1/2) \ln[{\rm det}G/{\rm det}{\tilde G}]}\right)
\label{jjjj}
\ee
In fact, this term can be absorbed into a proper
redefinition of the dilaton term.
This can be seen most intuitively
by observing that using world-sheet reparametrization invariance
one can concentrate the world-sheet curvature on a single point
on the world sheet $\sigma_0$: 
\be
    R^{(2)} \rightarrow 4\pi \chi \delta (\sigma - \sigma_0)
\label{singlepoint}
\ee
where $\chi$ is the Euler characteristic. 
Correspondingly,
the dilaton term in (\ref{pithetatilde}),
which so far remained unaffected by the change of variables, 
takes the form $S_{dil} = \Phi (r(\sigma_0))$.
Observing that
the dual metric is such that ${\rm det}{\tilde G} = 1/{\rm det}G$,
and evaluating $G$ at $r(\sigma_0)$, one concludes that the last
two terms of (\ref{final2}) encode a {\it shift in the dilaton} 
\be 
  \Phi \rightarrow \Phi + \frac{\chi}{2}{\rm ln}{\rm det} G
\label{duality2}
\ee
The point $\sigma_0$ on the world sheet is arbitrary due to
reparametrization invariance which is assumed valid; hence
the shift (\ref{duality2}) should be considered as a generic 
shift of the dilaton field at any world-sheet point,
just as required by the definition (\ref{duality}) 
of the duality transformation.

\section*{Acknowledgements}
These results were already reported by one of us (G.A.-C.) at
{\it Common Trends in Condensed Matter and High Energy Physics}
(Chia, Italy, 1-7 September 1997)
and {\it Quantum Aspects of Gauge Theories, Supersymmetry and Unification}, 
(Neuchatel, Switzerland, 18-23 September 1997);
our thanks go to the participants of those meetings for feed-back and encouragement.
We also gladly acknowledge discussions with H.~Dorn and I.~Kogan.
This work was supported in part by funds provided by the Foundation Blanceflor
Boncompagni-Ludovisi and  P.P.A.R.C..

%\newpage

\end{document}